# Structure of the Martian dayside magnetosphere: two types.

O.Vaisberg[1], S.Shuvalov[1], E.Dubinin[2], K.Kim[1]

1. Space Research Institute RAS, Moscow, Russia

2. Max-Planck-Institute for Solar System, Gottingen, Germany

1. **Abstract**

The comprehensive suite of scientific instruments with high temporal resolution measurements on the Mars Atmosphere and Volatile Evolution (MAVEN) spacecraft allowed studies of the structure and properties of the dayside magnetosphere of Mars. The dayside plasma envelope was found on first Mars orbiters (Vaisberg et al., 1976, Gringauz et al., 1976), but only the high temporal and mass resolutions of MAVEN instruments allowed investigations of the structure and processes of plasma of the dayside magnetosphere, the plasma layer which exists between shocked solar wind and ionosphere (Vaisberg et al., 2017). It was found that in the different dayside conditions the two different plasma layers form between the magnetosheath and ionosphere: (1) the mixture of heated ionospheric ions and energetic pick-up exospheric ions and (2) the layer of accelerated ionospheric ions population that forms the lower energy part of the plume formation.

The first type of dayside magnetosphere of Mars is formed by the interaction of pick-up oxygen ions of exospheric origin with the topside of Mars' ionosphere (Vaisberg and Shuvalov, 2021). The second plasma layer between magnetosheath flow and ionosphere forms from initial lower energy part of the plume that is accelerated and continues in the magnetosheath.

In this paper we show typical plasma populations, their properties and conditions that lead to formation of the dayside magnetosphere mainly from the material provided by the upper layer of the ionosphere. As the annex we include 115 dayside magnetosphere crossings of both types that were analyzed in the paper.

2. **Introduction**

The first close encounter of Mariner IV with Mars on 14-15 July 1964 at the minimum distance of 13,300 kilometers suggested that the solar wind interacts directly with the Martian atmosphere. The first crossings of the Martian magnetosphere were performed by Mars 2, Mars 3 and Mars-5 spacecraft in 1970 s and found the increase of the magnetic field up to ~30 nT (Dolginov et al., 1976) and the appearance of ions with lower energy than in the magnetosheath both on the dayside and well downstream (Bogdanov & Vaisberg, 1975). Unlike Earth, Mars lacks a global intrinsic magnetic field, which makes Mars plasma environment much smaller. In the early years of space exploration it was thought that the obstacle to the solar wind is formed by magnetic field from ionospheric currents induced by solar wind motional electric field (e.g. Dessler, 1968).

The term "magnetosphere of Mars" was first introduced with early Mars exploration (Van Allen et al., 1965) and was used for describing the Martian magnetotail, while the dayside magnetosphere remained unexplored for a long time. The region between magnetosheath and ionosphere was interpreted in terms of existence of the boundary layer on the dayside of Mars

(Szego et al., 1998). The Mars Express (MEX) spacecraft have measured the ion composition of plasma within this region (Dubinin et al., 2008a, 2008b). The magnetic pileup with a discontinuity (called the magnetic pileup boundary) was found, although in some cases a distinct pileup of the magnetic field was absent. A 'boundary' of the ionospheric photoelectrons was also observed, often accompanied by an abrupt increase (up to ~$10^3$ cm$^{-3}$) of the number density of the ionosphere plasma.

It was not until September 2014 that the Mars Atmosphere and Volatile EvolutioN (MAVEN) spacecraft carrying comprehensive plasma instrumentation arrived to Martian orbit and provided an opportunity to study this region in detail. There are several papers discussing the structure of the boundary and processes within it (Holmberg et al., 2018, Espley et al., 2018, Halekas et al., 2019). According to Halekas et al., 2017, "the magnetosphere of Mars is formed as a result of the direct and indirect interaction of the solar wind with the Martian ionosphere, through a combination of induction effects and mass loading". However, the dayside magneto-plasma envelope of Mars is far from sufficient understanding of structure and processes of the interaction between the shocked solar wind and the ionosphere. There is a multitude of nominations of the boundaries and envelopes in this part of Mars, not saying about processes going on in this part of Mars magnetosphere.

In this study we will consider two different types of the magnetosphere structure between the shocked solar wind flow and ionosphere. The first one is characterized by the existence of plasma that is a mixture of the pick-up exospheric ions and the heated and accelerated ionospheric ions (Vaisberg and Shuvalov, 2021). The second type is characterized by the existence of the layer with nearly mono-energetic ionospheric ions accelerated by solar wind electric field and called "ion plume" (Kallio and Koskinen, 1999; Kallio et al., 2006, 2008, Boesswetter et al., 2007, Dubinin et al, 2006, 2011, Liemohn et al., 2013, Dong et al. 2015). The low energy part of the plume is located below the boundary of the Martian magnetosphere. In this paper we analyze and compare the structure of both different types of the magnetosphere structure.

## 3. Instrumentation

The MAVEN spacecraft arrived at Mars in September 2014 to study the processes in the upper atmosphere/ionosphere and its interaction with solar wind and the escape of atmospheric species to space (Jakosky et al., 2015). MAVEN was inserted into an elliptical orbit with periapsis and apoapsis of approximately 150 km and 6200 km, respectively, and with a period of 4.5 hours.

In this paper we discuss observations mainly made by the Supra-Thermal And Thermal Ion Composition (STATIC) instrument from July to October 2019, during the solar minimum. The STATIC instrument mounted on the Actuated Payload Platform (APP) is used to study characteristics of different ion species at solar wind-Mars interaction. The instrument consists of a toroidal top hat electrostatic spectrometer with an electrostatic deflector at the entrance providing 360º x 90º field of view combined with a time-of-flight velocity analyzer resolving the major ion species ($H^+$, $He^{++}$, $He^+$, $O^+$, $O_2^+$, $CO_2^+$). It measures energy spectra of ions with different (m/q) in the range of 0.1 eV-30 keV with minimum cadence of 4 sec (McFadden et al., 2015). The measurements allow a retrieval of the velocity distribution functions and their moments (density, velocity, temperature). STATIC level 2 d1 data product was used for investigating ions of different species. This dataset contains differential energy fluxes for ions over 32 energy steps, 4 polar and 16 azimuth angles for 8 mass bins. The measurement cadence for the time intervals presented in the paper is 4 s. At high count rates some proton data can be wrongly registered as ions of bigger masses due to incorrect identification of start/stop signals of the time-of flight scheme. In order to diminish this effect, we applied a special procedure for $O^+$ and $O_2^+$ data, in which 8% of protons differential energy flux was subtracted from oxygen ion data for the same energy and angular bins.

Along with STATIC observations we used data obtained by the Solar Wind Ion Analyzer (SWIA, Halekas et al. 2015), a top-hat instrument with 360º x 90º field of view mounted on the solar array panel, Solar wind Electron Analyzer (SWEA, Mitchell et al., 2016), and by the measurements of the magnetic field with 32 Hz cadence (MAG, Connerney et al., 2015). For SWIA and SWEA measurements level 2 data products with survey spectra measurements were used, which provide the data with 4 and 2 seconds cadence, respectively.

## 4. Observations

This analysis is based on the observations made on the dayside of the Martian magnetosphere. The used data were selected by the following criteria: (1) during low solar minimum (having more non-disturbed cases for analysis), (2) dayside magnetosphere boundary crossings (objet of paper) (3) the regions without crustal magnetic field anomalies (having less additional influences) factors avoiding magnetic disturbances, implying crossings in the northern hemisphere. Altogether 115 crossings of magnetosphere boundary were considered within time interval from 07/27/2019 to 10/31/2019 for solar-zenith angles (SZA) from ~60° to ~ 95° (Fig. 1). Distribution of SZAs of the observed crossings is given in figure 1. This "landscape" of crossings was classified as: (a) 50 orbits with the ion plume originated within the magnetosphere and extended far to the magnetosheath, and (b) 65 orbits without a plume feature but with the observations of pick-up oxygen ions in the sheath/upper ionosphere and the cold ionospheric component within magnetosphere.

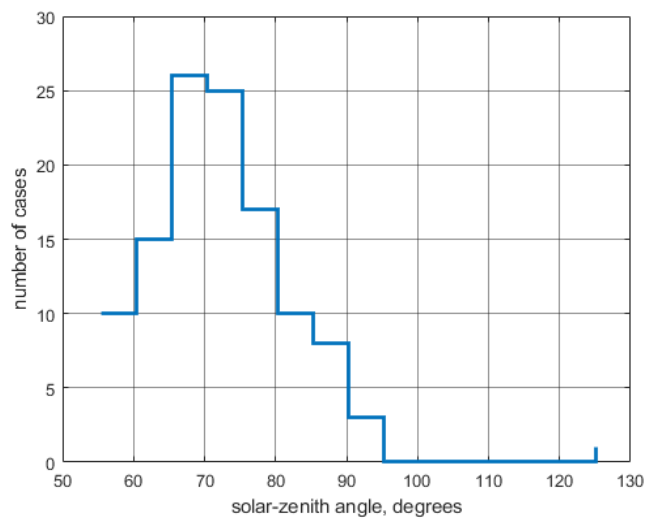

*Fig. 1. Distribution of the magnetosphere locations' the solar-zenith angles of observed cases. Solar-zenith angle for each crossing was calculated at the moment of magnetopause crossing, defined as $\frac{n_p}{n_p+n_h} \approx 0.5$ based on STATIC measurements.*

We present a detailed description of two typical crossings of magnetosheath-ionosphere region showing the properties of Martian magnetosphere with and without plume feature.

### 4.1 Dayside magnetosphere – 1st version

From 115 selected magnetosphere boundary crossings in 2019 this type of the plasma envelope was observed in about 60% of cases.

Figure 2 shows data obtained on July 29, 2019 (03:10.21 - 03:32.09 UT). The spacecraft moved from the magnetosheath into the magnetosphere. The SZA at magnetosphere boundary crossing was ~ 69.5°. There are several approaches for identification of different regions of plasma envelope and boundaries between them. The most often used method of the identification of the magnetosphere boundary is based on a drop of flux of the solar wind protons. Here we have two distinct changes in the proton spectra: at ~ 03:17:30 and at ~ 03:20 UT. The feature observed at ~ 03:17:30 UT is due to rotation of the platform with the STATIC spectrometer. At ~ 03:20 UT we observe a splitting of the proton spectra into two components: the low energy (100-200 eV) component and high-energy component. Protons with higher energy of the solar wind origin and picked up protons originated from the extended hydrogen atmosphere can penetrate deeper inside the magnetosphere due to their large Larmor radii. The origin of the low energy component is not so clear. It might consist of the sheath protons which gradually lose their momentum and protons of the atmospheric origin which gain the momentum from the solar wind. The position of the boundary at ~03:20 UT is confirmed by the measurements of the SWIA instrument which does not change the viewing direction. The boundary of the induced magnetosphere might be easily determined by a sharp change of the ion composition seen in Fig. 1 on the ion spectra and from the ratio $n_p/(n_p+n_h) \approx 0.5$, where $n_p$ is the proton number density and $n_p$ is number density of $O^+$ and $O_2^+$ ions. A decrease of the level of the magnetic field fluctuations observed at ~ 03:20 UT is in agreement with our identification of the boundary.

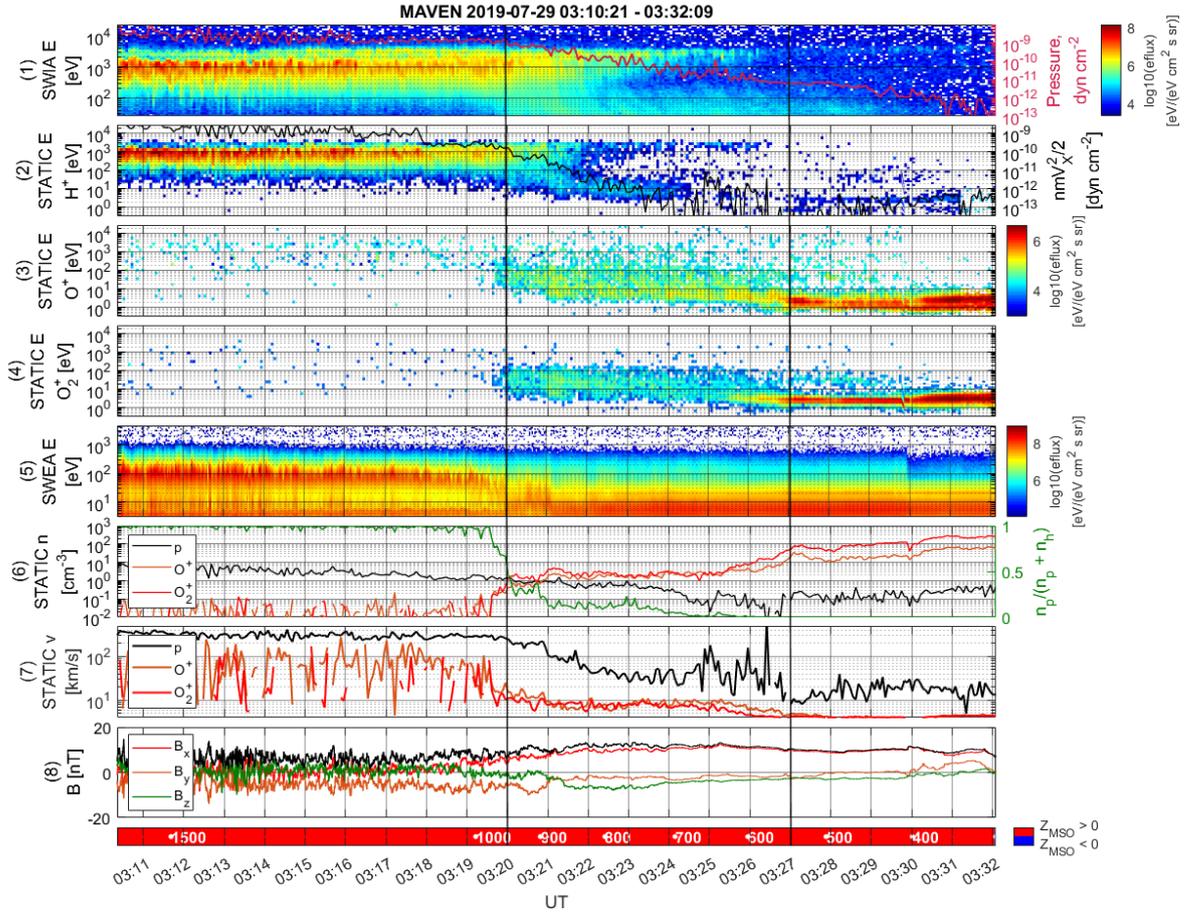

*Fig. 2. From top to bottom: (1) energy–time spectrogram summed over all ion species (protons, $O^+$ and $O_2^+$ ions); the red line depicts the ion dynamic pressure $n*m*V^2/2$, assuming that all measured ions are protons, (2-4) the energy-time spectrograms of protons, $O^+$ and $O_2^+$ ions with black line overlaying the proton momentum flux, (5) the energy-time spectrogram of electrons. (6) the number densities of protons, $O^+$, $O_2^+$ ions, and the ratio of $n_p$ to sum of the density of heavy ions $n_h$ ($O^+ + O_2^+$), (7) bulk velocities of different ion species, (8) three components and the magnitude of the magnetic field in MSO coordinate system. A drop at ~03:18 UT on the second panel is due to change of the STATIC viewing direction. The absence of pick-up ions with energies above ~ 500 eV after ~ 03:29.30 UT (note the glitches in diagrams of $O^+$ and $O_2^+$ at this time) is due to change of the measurement mode. The data in panels 2, 3, 4, 6, 7 are from STATIC, 1- SWIA, 5 - SWEA, 8 - MAG. From 112 selected magnetosphere boundary crossings in 2019 this type of the plasma envelope was observed in about 60% of cases. Two black vertical lines specify the location of magnetosphere.*

The third and fours panels of Fig.2 show the behavior of oxygen ions. We observe fluxes of $O^+$ ions with energies of about 100-5000 eV in the magnetosheath. These ions are originated from oxygen corona and accelerated by the motional solar wind electric field. They are also seen within the magnetosphere where a certain fraction of them have a higher energy. The $O^+$ ions with the energy below 100 eV and the ions $O_2^+$ with energies of ~ 5-330 eV are predominantly accelerated at closer distances and could not gain a higher energy.

The two boundaries within the transition from magnetosheath to ionosphere can be identified using plasma and magnetic parameters in Fig. 2. The outer boundary is quite sharp and is defined by number of physical parameters (going from magnetosheath to the magnetosphere): the drop of the energy flux of protons, the sharp appearance of pick-up ions $O^+$ and $O_2^+$ with energies 3-300 eV, drop of electrons energy, sharp drop of electron temperature, drop of ratio $n_p/(n_p+n_n)$. The shocked solar wind ion drops by the order of magnitude close to several other boundaries and defines the magnetopause. Some of these boundaries have been discussed by quite of few authors. Comparisons of different criteria used for identification of magnetosphere boundary (Holmberg et al., 2018, Espley et al., 2018, Halekas et al., 2019) show that the sharp drop by factor of 10 of the solar wind proton pressure is a better parameter and is often coincides with a sharp change in the composition.

The transition from the dayside magnetosphere to ionosphere is quite smooth and requires more analysis. The following parameters are taken into account: drop of ionospheric ions temperature, increase of their density, and the drop of pick-up ions number density. The following variations of $O_2^+$ and $O^+$ ions number densities and their ratio are observed in Fig. 1 in transition from magnetopause to ionosphere: (1) number densities of these ions jump from background to ~ 1-2 $cm^{-3}$ at 03:20:00 and continue to ~ 03:25:30 with small variations (2) within the time interval ~ 03:25:30 s - ~ 03:27:20 the number densities increase with $O_2^+$ faster than increase of $O^+$ and (3) the number densities of two ions show small increase with the ratio $O_2^+/O^+$ ~5. The spacecraft entered the ionosphere approximately at ~ 03:27:00 UT where the cold ion population dominates. Variations of $O_2^+/O^+$ ratio within magnetosphere will be discussed later as it provides the tool for analysis of dayside magnetosphere.

Fig.3 shows the angle between ion bulk velocity and the external normal to the surface of the planet for $O^+$ and $O_2^+$ ions. It is seen that ions in magnetosheath (before 03:20 UT) mainly move towards the planet as the above mentioned angle is > 90°. This happens because the particle drift velocity of ions picked-up in the oxygen corona is approximately co-directed with the solar wind velocity. The bulk velocity directions of both ion components below magnetopause remain stable at ~40° relative to the surface normal, indicating that they tend to move away from the planet.

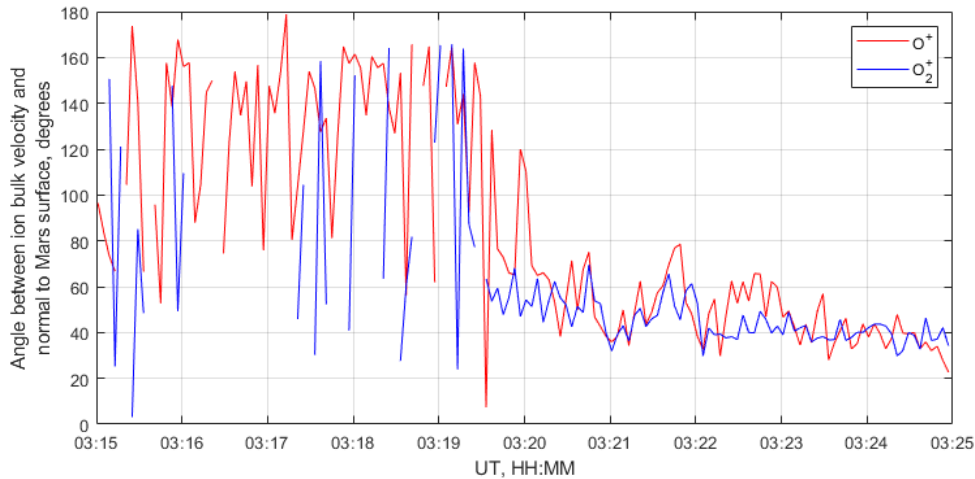

*Fig. 3. The angle between ion bulk velocity and normal to the surface of the planet for $O^+$ and $O_2^+$ ions for crossing presented in Fig. 2.*

Figure 4 shows a set of two-dimensional velocity distributions of $O^+$ ions in the MSE coordinates, in which the X-axis is pointing to the Sun, the Y-axis is along the cross-flow component of the interplanetary magnetic field in the solar wind, and the Z-axis is along the motional electric field $\mathbf{E} = -(1/c)\,\mathbf{v} \times \mathbf{B}$, for about the same time interval as in Fig. 2. The size of the circles shows a phase space density of ions in the velocity space. It is observed that with increase of the distance from the ionosphere, Z-component of the velocities increases indicating ion acceleration by the motional electric field.

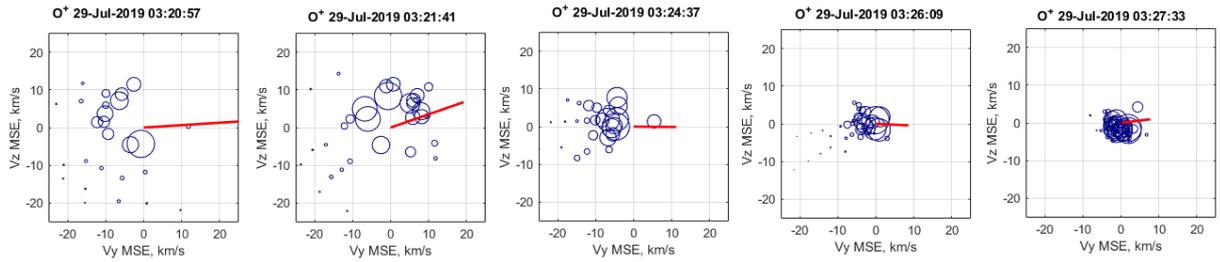

Fig. 4. Examples of the distribution functions of oxygen ions in the velocity (Vz-Vy) space. The size of the circle is proportional to the phase space density of ions. The red lines show the magnetic field projection onto the YZ-MSE plane.

The time interval ~03:25 - ~03:27 approximately corresponds to the transition from the lower part of the ionosphere where the ratio $O_2^+/O^+$ is about of ~ 2 to the topside ionosphere with $n(O_2^+)/n(O^+)$ ~ 1 (see Fig. 2). At higher altitudes we observe a broadening of the ion distributions in the direction of the motional electric field (03:24:27) and further gradual ion acceleration (03:21:41-03:20:57 UT).

A significant heating of oxygen ions in the outer part of the plasma envelope is clearly seen in Fig. 5 which shows the energy spectra of the protons, $O^+$ and $O_2^+$ ions. It is seen that with approach to the ionosphere the ion spectra become less energetic. In the ionosphere (~03:27:20) oxygen ions become colder.

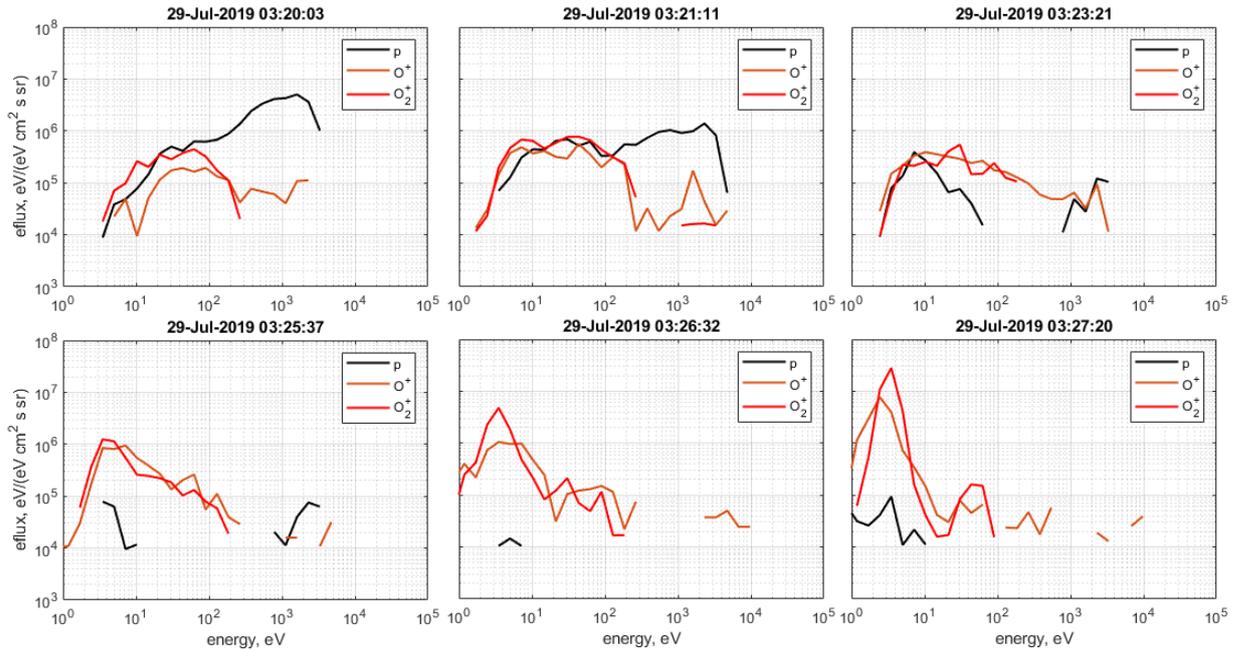

Fig. 5. Energy spectra of $H^+$, $O^+$ and $O_2^+$ ions in the plasma envelope adjacent to the ionosphere.

Figure 6 shows another example of dayside magnetosphere at 2019-07 31. Magnetosheath is shown from the left at ~800 km. Criteria for locations of the boundaries were presented in discussion of previous case in Fig. 1. In the case shown in Fig. 5 the magnetopause was crossed at 06:32:00 UT at altitude ~620 km that is indicated by the ratio $n_p/(n_p+n_n) \approx 0.5$, the sharp drops of ions energy (panel 1) and protons energy (panel 2), and by rise number densities and the temperatures of ions $O^+$ and $O_2^+$ fluxes (panels 3 and 4 ). Ionopause is identified at altitude ~ 440 km on ~ 06:35.40 UT, by the drop of the hot ions.

Another evidence of the transition from magnetosphere to ionosphere is the increase of $O_2^+/O^+$ number densities ratio of from 1-2 (panel 6) before 06:35.40 UT to ~ 5-10 after that time. This type of $O_2^+/O^+$ number density variation along transition from magnetopause to ionopause is quite similar to one in Fig. 2, namely this ratio remains ~1-2 while increasing magnitudes of number densities two ion species by about by factor of 10 with increase of $n_{O2+}/n_{O+}$ ratio to 10 at the cold ionospheric ions.

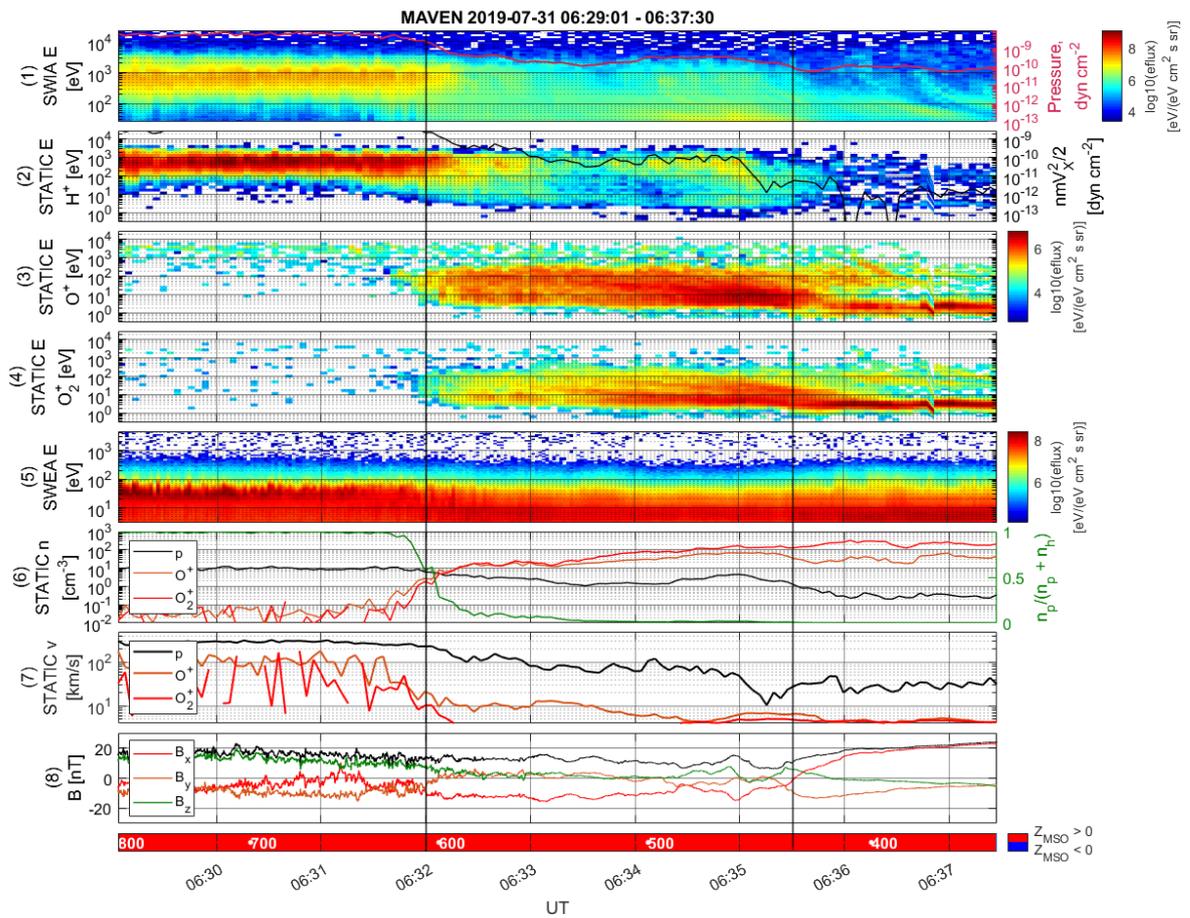

*Fig. 6. See caption in Fig. 2.*

Figure 7 shows the directions of ions within magnetosheath and within magnetosphere. It is seen that ions in magnetosheath (before 03:20 UT) predominantly move towards the planet as the above mentioned angle is > 90°. The bulk velocity directions of both ion components below magnetopause remain stable at ~80° relative to the surface normal, indicating that they move away from the planet.

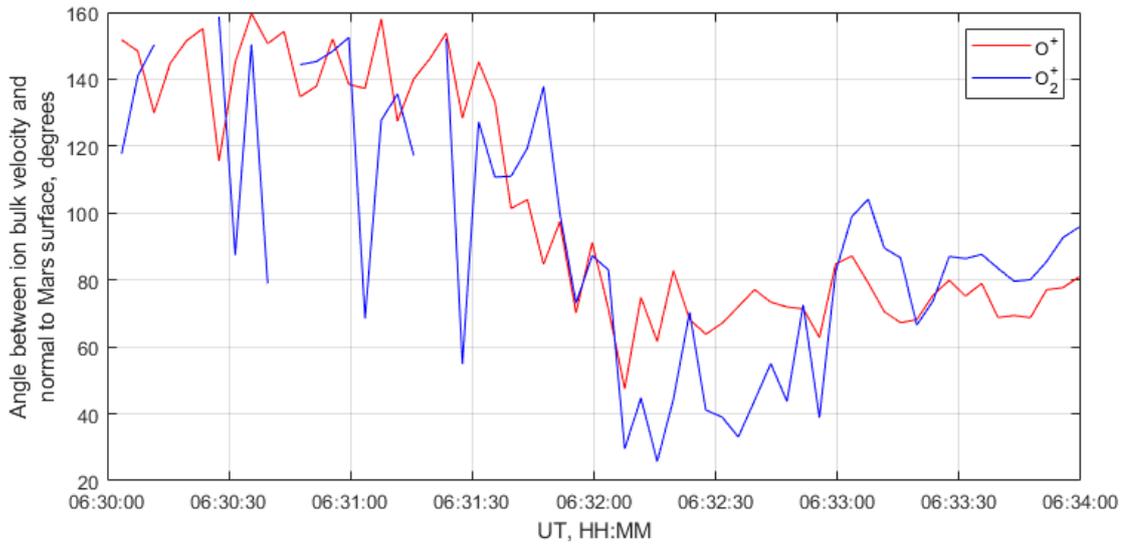

*Fig. 7. The angle between ion bulk velocity and normal to the surface of the planet.*

**4.2 Dayside magnetosphere with a plume feature**

When the ion plume is present in the dayside magnetosheath, the plasma in the magnetosphere also contains populations of accelerated ionospheric ions and pick-up ions. The characteristics of ionospheric and pickup ions are quite similar to those observed within the dayside magnetosphere in cases without a plume feature. Figure 8 shows the typical example of such magnetosphere structure.

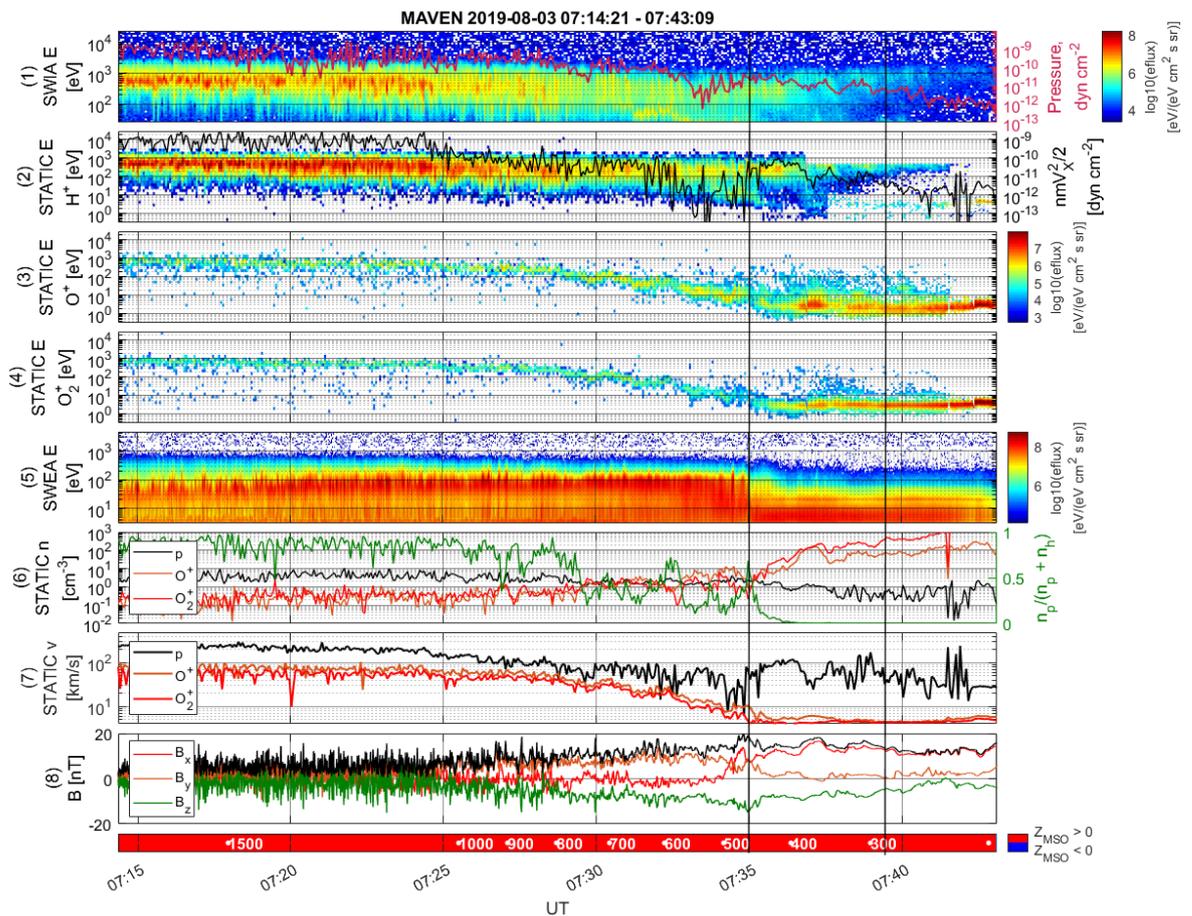

*Fig. 8. The same parameters as in Figure 2 are shown here for the orbit on Aug. 3, 2019. The ion plume is characterized by a narrow energy spread of the heavy ions with monotonic energy increase from ionosphere through the magnetosheath.*

The plume is observed on 2019-08-03 UT at 07:14.21 - 07:43.09 UT (Fig. 8). It includes a significant part of magnetosheath (~07:14 UT - ~07:32.30 UT) with a drop of ion the dynamic pressure from $1 \times 10^{-9}$ dyn/cm$^2$ by a factor of 10 which we identify as the boundary. A further steep increase of heavy ions number density and $n_p/(n_p+n_h)$ ratio (6th panel), increase in magnetic field magnitude B, and decrease of magnetic field fluctuations (8th panel) confirm our identification of this boundary. The SZA at the magnetopause crossing was ~ 75°.

Panels 3 and 4 show that magnetosheath is filled with ions of the plume. This plasma structure characterized by narrow energy spectra of the heavy ions with monotonic increase in energy is formed by ions originated in the extended exosphere and the upper ionosphere by the motional solar wind electric field.

The angle between heavy ions bulk velocity and normal to the surface of the planet presented in figure 9 shows that these ions move away from the planet. This angle decreases as ions accelerate in the motional electric field. A noticeable feature is that the angle differs by ~10° for both ion species. It may be related with the different gyroradii of these ions.

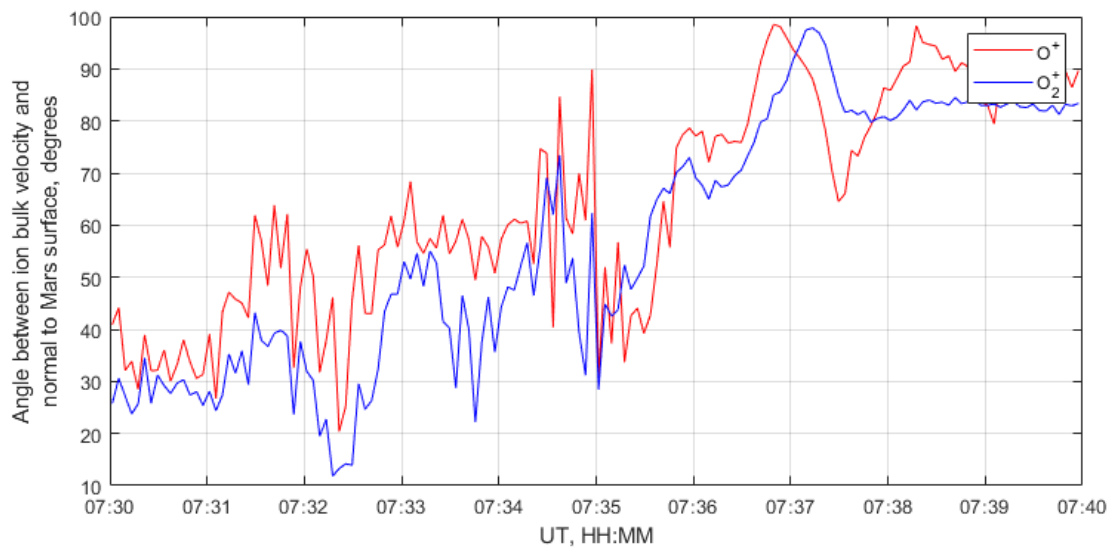

*Fig. 9. The angle between ion bulk velocity and normal to the surface of the planet for $O^+$ and $O_2^+$ ions for crossing presented in Fig. 8.*

Figure 10 shows the energy distributions of $O^+$ and $O_2^+$ ions along the curve of plume ions. These energy distributions very much differ from pick-up ions distributions (Fig. 5) indicating the approximate unchanged velocity distribution along the plume curve.

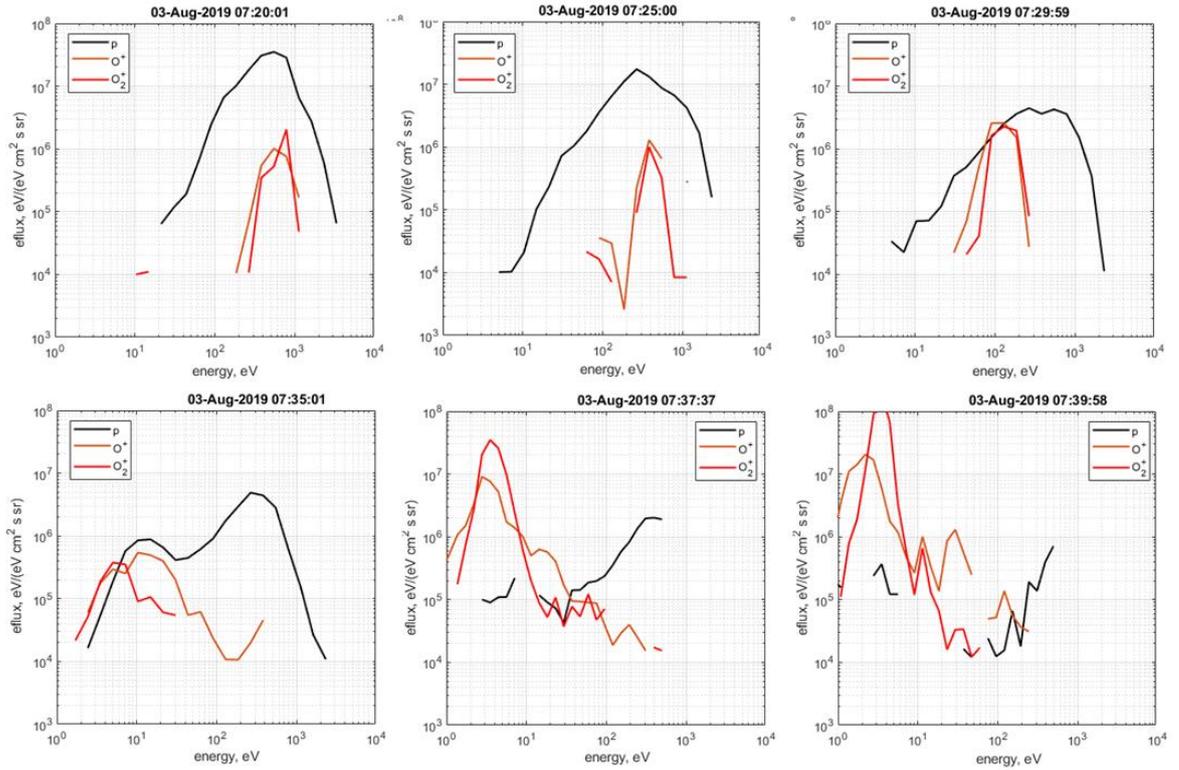

*Fig. 10. Protons and heavy ions energy spectra in different regions.*

Figure 10 shows selected energy spectra of protons and planetary ions in the magnetosheath and ionosphere. Note very narrow energy spectra of planetary ions in the plume.

We also observe fluxes of the $O^+$ and $O_2^+$ pick-up ions with energies of $10eV$-$10^3$ eV in the magnetosheath. Their densities in magnetosheath are rather small (~0.1 $cm^{-3}$), however, they increase monotonically as the s/c move towards the ionosphere and reach ~20 $cm^{-3}$ in its upper part.

Similarly, this ion component gradually decays during the transition from heated topside ionosphere to a regular ionosphere at ~07:42 UT which indicate that the plume originate from the ionosphere.

Another example of the magnetosphere when plume is observed is shown in Fig. 11.

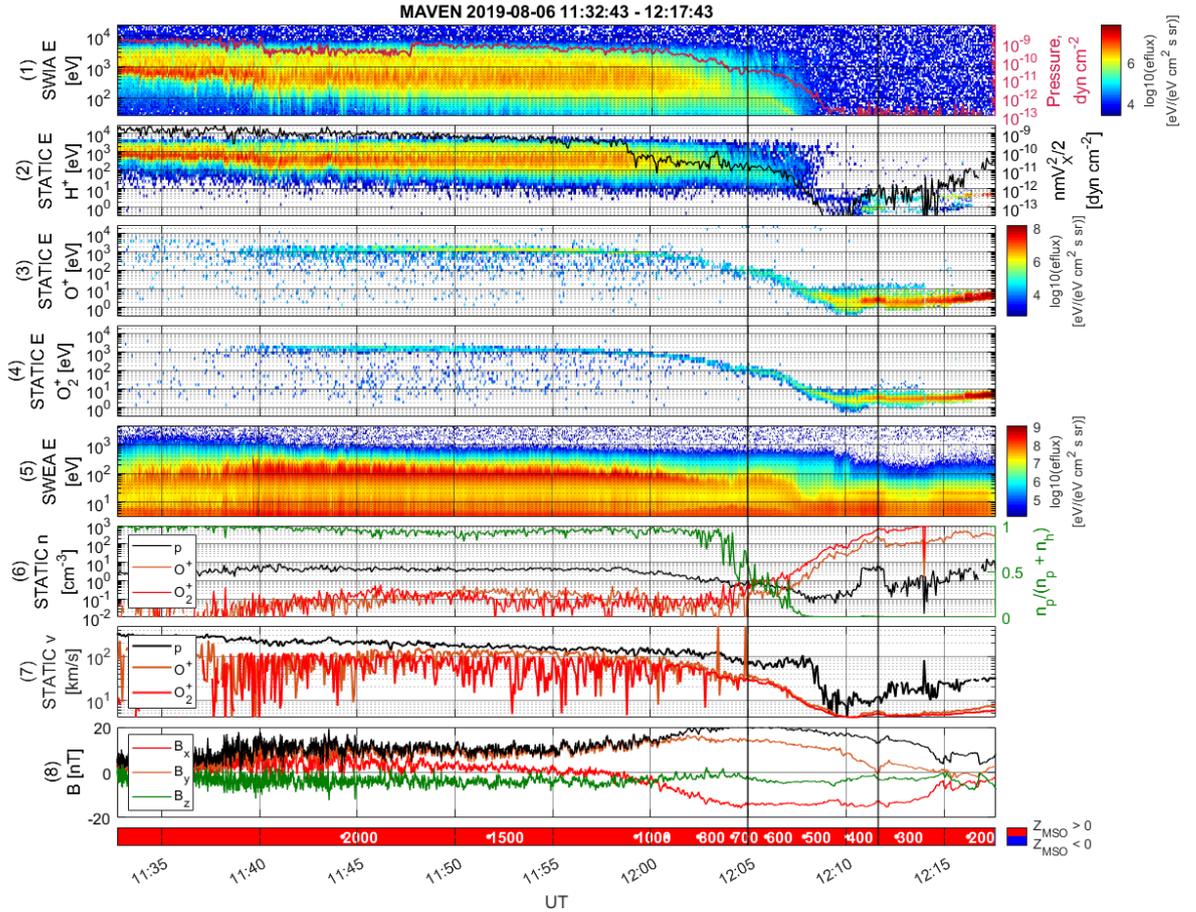

*Fig. 11. Another example of dayside magnetosphere in the case of plume. See caption in Fig. 2.*

Using the ratio $n_p/(n_p+n_n) \approx 0.5$ one puts the boundary at 12:05:00 UT. This time agrees with the drop of total magnetosheath ions dynamic pressure by factor of 10. This estimation indicate that the ions energies below of ~ $10^2$ energy fluxes of $O^+$ and $O_2^+$ ions are accelerated within magnetosphere.

The ions $O_2^+/O^+$ number densities ratios from smaller value of ~3 in the magnetosphere increases to ~ 8 in the ionosphere. It appears that the ratio of $O_2^+/O^+$ in the upper ionosphere is determined by UV ionization and atmospheric processes, the same ratio value within magnetosphere is determined mainly by UV ionization.

Like in the previous case of magnetosphere crossing with plume feature, bulk velocities of accelerated ions are directed away from the planet (see figure 12), and the angle between this direction and the planet surface normal decreases as particles accelerate.

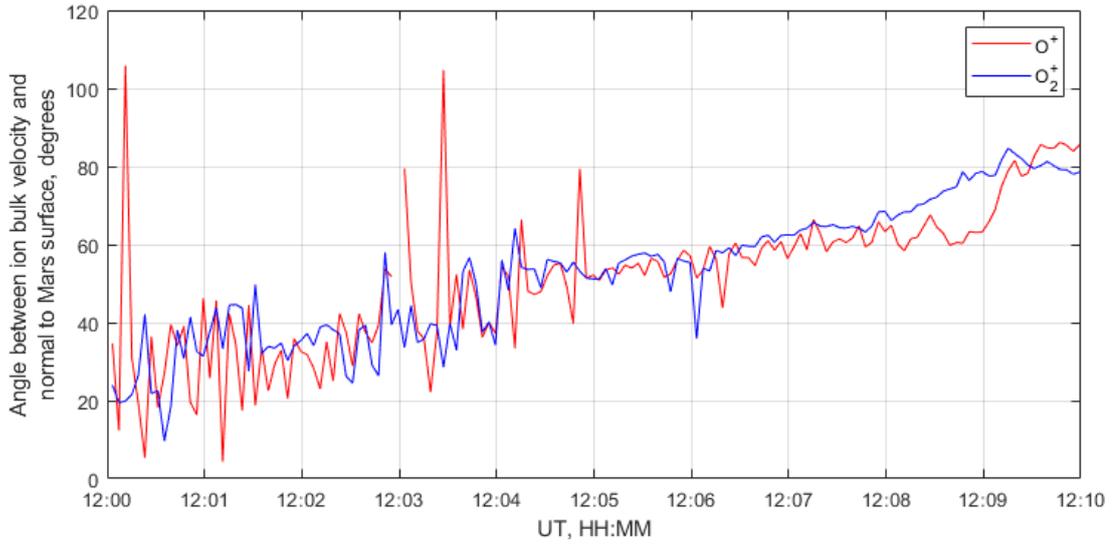

*Fig. 12. Same as figure 9 for crossing presented in figure 11.*

Figure 13 and Table 1 show the average ratios of $n_{O2+}/n_{O+}$ over 3 observed sets of magnetospheres crossings considered in this paper. The scatter of numbers is significant the average values are distinct.

Table 1. The ratio between number densities of $O_2^+$ and $O^+$ ions in different regions.

|  | $n_{O_2^+}/n_{O^+}$ in magnetopause | $n_{O_2^+}/n_{O^+}$ in magnetosphere | $n_{O_2^+}/n_{O^+}$ in ionosphere |
|---|---|---|---|
| mean value | 1.8 | 2.4 | 5.8 |
| standard deviation | 1.7 | 1.5 | 7.0 |

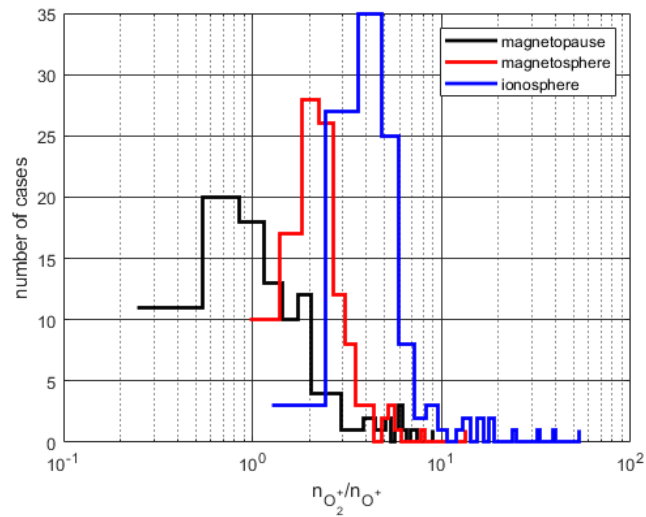

*Fig. 13. Distributions of $n_{O2+}/n_{O+}$ in observed crossings at magnetopause, magnetosphere and upper ionosphere.*

It is known that plume ions originate from ionosphere and are accelerated by motional electric field of the solar wind plasma (Dubinin et al, 2006 and references therein) making an important atmospheric loss channel (Dong et al., 2015). In order to make a rough estimation of the height from which the electric field starts to accelerate plume ions to the observed energy values we compared the measured energies in the plume with electric field potential drop between the spacecraft position and height of 420 km above the surface of Mars in the +$Z_{MSE}$ direction (Figure 14) for the crossing presented in figure 8. This height of 420 km was selected for best fitting between the two curves in Figure 8 and is also close to the upper boundary of the ionosphere. The blue line shows the energy value with the highest measured differential energy flux at a certain moment of time. The motional electric field was calculated from the measurements upstream from the Martian bow shock as **E** = - (1/c)[**V**$_{sw}$ × **B**], where **V**$_{sw}$ is the measured by the STATIC proton bulk velocity averaged over the time interval from 7:00 to 7:10 UT, and B is magnetic field vector measured by MAG and averaged over the same time interval.

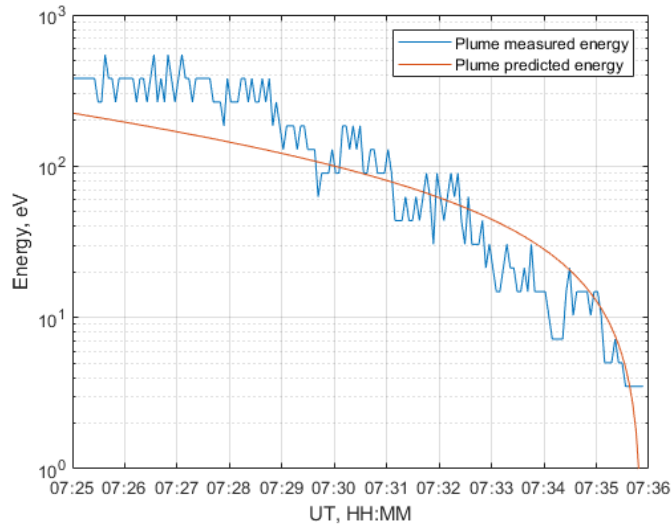

*Figure 14. Comparison between the energy in the plume (blue) and predicted energy in the assumption that plume ions are accelerated by induced solar wind electric field from 420 km height with zero initial energy (red).*

In general, the curves are in a good agreement with each other. Discrepancy between the curves seen in energies higher than 100 eV are probably caused by cycloidal motion of the plume ions, as a result of that the ions are detected further from spacecraft than predicted by our simple model and, accordingly, cover a longer path gaining an additional energy. It is seen that ions in the model are significantly under-accelerated in the left part of the figure, compared with the measured ion energy. It may be apparently related with the electric field in the magnetosheath that has not been taken into account and that is stronger than the upstream electric field. The proposed model can also be used for estimation of the height from which the electric field starts to accelerate ionospheric ions.

## 5. Summary and Conclusion

This analysis of plasma and magnetic field are bases on the MAVEN data. The measurements of plasma and magnetic field at 115 passes at altitude above ~300 km in the north hemisphere at solar-zenith angles ~70° within time interval 27.07.2019-31.10.2019 were analyzed. Previous analysis of other dayside sets of regions between the dayside magnetosheath and ionosphere did not include the cases with the plume. In current selection of passes of dayside magnetosphere we included the plume cases and it was found that the ratio of cases without the plume (cases A) to the cases with plume (cases B) was approximately 60% to 40%.

Previous analysis (Vaisberg and Shuvalov 2021) did not found the cases of direct interaction of the dayside magnetosheath flow with the ionosphere. The layer between magnetosheath and ionosphere was filled with the mixture of the heated $O^+$ and $O_2^+$ ions and pick-up and called dayside magnetosphere of Mars

More detailed analysis of magnetosphere plasmas showed quite regular profiles (A) of $O^+$ and $O_2^+$ number densities and the ratio of these ions (see Figure 2 and Figure 6). (1) initial steep increase at magnetopause with $O_2^+/O^+ \sim 1$, (2) smooth or two steps increase by factor of 1.5-3 with variations $O_2^+/O^+ \sim 1-2$, and (3) short increase of $O^+$ especially $O_2^+$ significant increase of $O_2^+/O^+$ ratio, an (4) flat or slow rising of $O_2^+$ and $O^+$ with values specific to upper ionosphere.

In cases B (the plume, Fig. 8 and 11) one can see the layer of heated ionospheric layer above cold ionosphere: average energy increase up to ~ 10 eV ions in Fig. 8 and up to ~$10^2$ eV in Fig. 11. These are the boundaries correspond to the location of obstacles which correspond to the drops of the magnetosheath energy flux for both of the cases. Figures 9 and 12 confirm the locations of boundaries between magnetosheath flows and the obstacles.

In the summary, the plume starting on the dayside of Mars participates in the formation of the region that is the same role as the dayside magnetosphere, being an obstacle between magnetosheath and ionosphere.

## Acknowledgements

The work was supported by the Russian Science Foundation by grant 21-42-04404. MAVEN data are publicly available through the Planetary Data System (https://pds.nasa.gov). ED, who made an important contribution to sections 2, 5, 6, wishes to acknowledge support from DFG for supporting this work by grants TE 664/4-1 and PA 525/25-1.